\documentclass[apsscuseria06,prl,preprint,showpacs,amsmath,floatfix,superscriptaddress]{revtex4}

\usepackage{graphicx}

\begin{document}

\title{Localization of Metal-Induced Gap States at the Metal-Insulator
  Interface:\\Origin of Flux Noise in SQUIDs and Superconducting Qubits}

\author{SangKook Choi}
\affiliation{Department of Physics, University of California, Berkeley, California 94720} \affiliation{Materials Sciences
Division, Lawrence Berkeley National Laboratory, Berkeley,
California 94720}

\author{Dung-Hai Lee}
\affiliation{Department of Physics, University of California, Berkeley, California 94720} \affiliation{Materials Sciences
Division, Lawrence Berkeley National Laboratory, Berkeley,
California 94720}

\author{Steven G. Louie}
\affiliation{Department of Physics, University of California, Berkeley, California 94720} \affiliation{Materials Sciences
Division, Lawrence Berkeley National Laboratory, Berkeley,
California 94720}

\author{John Clarke}
\email{jclarke@berkeley.edu}
\affiliation{Department of Physics, University of California, Berkeley, California 94720} \affiliation{Materials Sciences
Division, Lawrence Berkeley National Laboratory, Berkeley,
California 94720}

\date{\today}

\begin{abstract}
The origin of magnetic flux noise in Superconducting Quantum
Interference Devices with a power spectrum scaling as $1/f$ ($f$ is
frequency) has been a puzzle for over 20 years. This noise limits the
decoherence time of superconducting qubits. A consensus has emerged that
the noise arises from fluctuating spins of localized electrons with an
areal density of $5\times10^{17}$m$^{-2}$. We show that, in the presence of
potential disorder at the metal-insulator interface, some of the
metal-induced gap states become localized and produce local moments. A
modest level of disorder yields the observed areal density.
\end{abstract}


\maketitle

Well below 1 K, low-transition temperature Superconducting Quantum
Interference Devices \cite{clarke04} (SQUIDs) exhibit magnetic flux
noise \cite{clarke87} with a temperature-independent spectral density
scaling as $1/f^\alpha$, where $f$ is frequency and 0.6
$\leq \alpha \leq 1$.
The noise magnitude, a few
$ \mu\Phi_0$Hz$^{-1/2}$
at 1
Hz ($\Phi_0$ is the flux quantum), scales slowly with the SQUID area, and does
not depend significantly on the nature of the thin film superconductor or
the substrate on which it is deposited. The substrate is typically
silicon or sapphire, which are insulators at low temperature ($T$) \cite{clarke87}. Flux noise of similar magnitude is observed
in flux \cite{yoshihara06,kakuyanagi07} and
phase \cite{bialczak07} qubits. Flux noise limits the
decoherence time of superconducting, flux sensitive qubits making
scale-up for quantum computing problematic. The near-insensitivity of
noise magnitude to device area \cite{clarke87,bialczak07,lanting09} suggests the origin of the noise is
local. Koch \textit{et al.} \cite{clarke07} proposed a model in which electrons hop
stochastically between traps with different preferential spin
orientations. A broad distribution of time constants is necessary to
produce a $1/f$ power spectrum \cite{machlup54,dutta81}. They found that the major noise
contribution arises from electrons above and below the superconducting
loop of the SQUID or qubit \cite{bialczak07,clarke07}, and that an areal density of about $5 \times
10^{17}$m$^{-2}$ unpaired spins is required to account for the observed noise
magnitude. De Sousa \cite{desousa07} proposed that the noise arises from spin flips of
paramagnetic dangling bonds at the Si-SiO$_2$ interface. Assuming an array
of localized electrons, Faoro and Ioffe \cite{faoro08} suggested that the noise
results from electron spin diffusion. Sendelbach \textit{et al.} \cite{sendelbach08} showed that
thin-film SQUIDs are paramagnetic, with a Curie ($1/T$) susceptibility. Assuming the paramagnetic moments arise from localized
electrons, they deduced an areal density of $5 \times 10^{17}$m$^{-2}$. Subsequently,
Bluhm \textit{et al.} \cite{bluhm_ar} used a scanning SQUID microscope to measure the
low-$T$ paramagnetic response of (nonsuperconducting) Au rings
deposited on Si substrates, and reported an areal density of $4\times10^{17}
\mathrm{m}^{-2}$ for localized electrons. Paramagnetism was not observed on the bare
Si substrate.

In this Letter we propose that the local magnetic moments originate in
metal-induced gap states (MIGS) \cite{cohen76} localized by potential disorder at the
metal-insulator interface. At an ideal interface, MIGS are states in the
band gap that are evanescent in the insulator and extended in the
metal \cite{cohen76} (Fig.1). In reality, at a nonepitaxial metal-insulator
interface there are inevitably random fluctuations in the electronic
potential. The MIGS are particularly sensitive to these potential
fluctuations, and a significant fraction of them--with single
occupancy--becomes strongly localized near the interface, producing the
observed paramagnetic spins. Fluctuations \cite{nyquist28} of these local moments yield
$T$-independent $1/f$ flux noise.

To illustrate the effects of potential fluctuations on the MIGS we start
with a tight-binding model for the metal-insulator interface, consisting
of the (100) face of a simple-cubic metal epitaxially joined to the
(100) face of an insulator in a CsCl structure (Fig. 2(a)). For the metal
we assume a single s-orbital per unit cell and nearest neighbor (NN)
hopping. For the insulator we place an s-orbital on each of the two
basis sites of the CsCl structure and assume both NN and next-nearest
neighbor (NNN) hopping. The parameters are chosen so that the metal
s-orbitals are at zero energy and connected by a NN hopping energy of
-0.83 eV. The onsite energy of the orbitals on the Cs and Cl sites is
taken to be -4 eV and 2 eV, respectively, and both the NN and NNN
hopping energies are set to -0.5 eV. These parameters yield a band width
of 10 eV for the metal, and 8 and 4 eV band widths, respectively, for
the valence and conduction bands of the insulator with a band gap of 2 eV
(Fig. 2(d)). These band structure values are typical for conventional
metals and for semiconductors and insulators. For the interface we take
the hopping energy between the metallic and insulating atoms closest to
the interface to be -0.67 eV, the arithmetic mean of -0.83 and -0.5 eV.

The electronic structure of the ideal metal-insulator junction is
calculated using a supercell \cite{louie75} containing $20\times20\times20$ metal unit cells and
$20\times20\times20$ insulator unit cells, a total of 24,000 atoms. The total
density of states (DOS) of the supercell (Fig. 2(e)) shows a nearly flat
DOS in the band gap region. The states in the insulator band gap are
MIGS that are extended in the metal, decaying rapidly away from the
interface into the insulator. Our model with a lattice constant of 0.15
nm yields an areal density of states for the MIGS of about
$3\times10^{18}$eV$^{-1}$ m$^{-2}$, consistent with earlier self-consistent
pseudopotential calculations \cite{cohen77}.

To mimic the effects of interfacial randomness, we allow the onsite
energy to fluctuate for both metal and insulator atoms near the
interface \cite{anderson58}. Specifically we assume an energy distribution $P(E) =
(1/\sqrt{2\pi}\delta)exp[-(E-E_0)^2/2\delta^2]$, where $E_0$ is the original onsite energy
without disorder, and $\delta$ is the standard deviation. We characterize the
degree of disorder by the dimensionless ratio $R = 2\delta/W$, where $W$ is the
bandwidth of the metal. For those MIGS that become localized, the energy
cost, $U_i$, for double occupation is large, and we cannot use a
noninteracting electron approach. Instead we adopt a strategy similar to
that used by Anderson in his calculation of local moment formation \cite{anderson75}. We
separate the space near the interface into 3 regions: (i) the perfect
metal region (M), (ii) an interfacial region consisting of 2 layers of
metal unit cells and 2 layers of insulator unit cells (D) (Fig. 2(b)),
and (iii) the perfect insulator region (I). Region (ii) is analogous to
the impurity in Anderson's analysis. 

We first compute the single-particle eigenstates,
$\varphi_i(\mathbf{r})$,
 of region D \emph{in
isolation}. For each of these states, we compute $U_i$ (using a long-range
Coulomb potential with an onsite cutoff of 10 eV) and the hybridization
energy $\mathit{\Gamma_i}$ due to hopping to the metal and the insulator
\cite{epaps}. With the
computed values of $U_i$ and $\mathit{\Gamma_i}$, we solve Anderson's equation for the
spin-dependent occupation for each localized state $\mid i\rangle$:
\begin{equation}\label{eq1}
\langle n_{i,\sigma}\rangle=\frac{1}{\pi}\int_{-\infty}^{E_F}dE'\frac{\mathit{\Gamma_i}}{(E'-E_{i,\sigma})^2+\mathit{\Gamma_i}^2}.
\end{equation}
Here, $E_{i,\sigma}=E_i+U_i\langle n_{i,-\sigma}\rangle$ and $\sigma$ is the spin index. The net moment associated with the state
is given by $m_i=\mu_\mathrm{B} | \langle n_{i,\sigma} \rangle - \langle
n_{i,-\sigma} \rangle|$. Equation (\ref{eq1}) and the associated expression for the net
moment of the localized states are calculated within the self-consistent
Hartree-Fock approximation \cite{anderson75}. An $m_i\ne0$ solution is obtained only when
$U_i/(E_F-E_i)$ exceeds a critical value which depends on $\mathit{\Gamma_i}/(E_F-E_i)$. In the
large $U_i$ limit, it is more appropriate to start from the weak coupling
limit ($\mathit{\Gamma_i}=0$), where the localized state is populated by a single
electron, and treat $\mathit{\Gamma_i}$ as a perturbation. By calculating the areal
density of such moment-bearing localized states we estimate the density of spin-$\frac{1}{2}$ local moments. 

Figure 3 shows the calculated distribution $\rho(E,U)$ in the isolated
interfacial region for $R$= 0.05, 0.1, 0.15, 0.2, 0.25, and 0.3; for each
value, higher values of $U$ correspond to more localized states. As
expected we see that, for any given degree of randomness, the states
with energy inside the insulator band gap (the MIGS) or those at the
band edges are most susceptible to localization. Figure 4 shows a
perspective plot of the charge density of two states, with high and low
values of $U_i$, showing the correlation between the degree of wavefunction
localization and the value of $U_i$. Both states are centered in the
insulator, a general characteristic of localized states in the band gap
originating from the MIGS.

Setting the Fermi energy at the insulator midgap value, we estimate the
areal density of spins for a given degree of randomness $R$. The top panel
in Fig. 5 depicts the distribution $\rho(E,m)$ of the spin moments as a
function of energy. We see that for small $R$ virtually all the local
moments are derived from the MIGS. The bottom panel of Fig. 5 shows the
calculated areal density of local moments versus $R$. Our simple model
thus indicates that moderate potential fluctuations ($R \sim 0.15$) at
the interface produce an areal density of localized moments comparable
to experimental values \cite{screening}. Although our analysis is for a specific model,
we expect the general physical picture to remain valid for real
materials. First, the formation of MIGS at a metal-insulator interface
is universal, and their areal density is rather insensitive to the
nature of the materials as discussed in supplements \cite{epaps} and shown
numerically in Ref. \cite{cohen77}. Second, the formation of local moments from the
combination of localized states and Coulomb interaction is a general
phenomenon \cite{anderson75}. We also note that our analysis should not be significantly
modified when the metal is superconducting. This is because the $U_i$ for
the localized states is generally much greater than the pairing gap. Of
course, extended states with negligible $U_i$ would be paired.

Given our picture of the origin of the localized spin-$\frac{1}{2}$ moments, how do
they produce 1/$f$ flux noise with a spectral density $S_\Phi(f)
\propto 1/f^\alpha$? The local moments interact via mechanisms such as direct
superexchange and the RKKY interaction \cite{faoro08,kittel54,kasuya56,yosida57} between themselves, and
Kondo exchange with the quasiparticles in the superconductor. This
system can exhibit a spin-glass transition \cite{weissman93}, which could account for
the observed susceptibility cusp \cite{sendelbach08} near 55 mK. For $T
> 55$ mK, however, experiments suggest that the spins are in thermal
equilibrium \cite{harris08} and exhibit a $1/T$ (Curie Law) static
susceptibility \cite{sendelbach08,bluhm_ar}. In this temperature regime, for $hf << k_BT$ standard
linear response theory \cite{fetter71} shows that the imaginary part of the dynamical
susceptibility $\chi''(f,T)=A(f,T)(hf/k_BT)$. Here, $A(f,T)\propto\sum_\mu\sum_{\alpha,\beta}P_\alpha\delta(hf+E_\alpha-E_\beta)|\langle\beta\mid
S_\mu\mid\alpha\rangle|^2$, where $S_\mu$ is the $\mu$-th component of
the spin operator, $\alpha$ and $\beta$ label the exact
eigenstates, and $P_\alpha$ is the Boltzmann distribution associated with state
$\alpha$. Combining the above result with the fluctuation-dissipation theorem \cite{nyquist28}
which relates the flux noise to $\chi''(f,T)$, namely
$S_\Phi(f,T)\propto (k_BT/hf)\chi''(f,T)$, we conclude that the observed
$1/f^\alpha$ spectral density implies $A(f,T) \propto 1/f^\alpha (0.6
\leq \alpha \leq 1)$. Assuming low frequency contributions dominate the
Kramers-Kronig transform, this result is consistent with the observed
1/$T$ static susceptibility, and the recent measurement
\cite{sendelbach_p} showing that
flux noise in a SQUID is highly correlated with fluctuations in its
inductance, However, without knowing the form of the interaction
between the spins, one cannot derive this behavior for $A(f,T)$ theoretically.

In conclusion, we have presented a theory for the origin of the
localized magnetic moments which have been shown experimentally to give
rise to the ubiquitous low-$T$ flux 1/$f$ noise observed in SQUIDs
and superconducting qubits. In particular we have shown that for a
\emph{generic} metal-insulator interface, disorder localizes a substantial
fraction of the metal-induced gap states (MIGS), causing them to bear
local moments. Although MIGS have been known to exist at metal-insulator
interfaces for three decades, we believe this is the first understanding
of their nature in the presence of strong local correlation and
disorder. Provided $T$ is above any possible spin glass
transition, experiments show that fluctuations of these local moments
produce a paramagnetic $\chi'$ and a power-law, $f$-dependent $\chi''$ which
in turn leads to flux 1/$f$ noise. It is important to realize that
localized MIGS occur not only at the metal-substrate interface but also
at the interface between the metal and the oxide that inevitably forms
on the surface of superconducting films such as aluminum and
niobium. There are a number of open problems, for example, the precise
interaction between the local moments, its relation to the value of $\alpha$,
and the possibility of a spin glass phase at low temperature. A particularly intriguing experimental issue to address is why different
metals and substrates evidently have such similar values of R, around
0.15. Experimentally, to improve the performance of SQUIDs and
superconducting qubits we need to understand how to control and reduce
the disorder at metal-insulator interfaces, for example, by growing the
superconductor epitaxially on its substrate.

We thank R.McDermott and K.A.Moler for prepublication copies of
their papers. S.C. and S.G.L. thank M.Jain and J.D.Sau for fruitful discussions. This
work was supported by the Director, Office of Science, Office of Basic
Energy Sciences, Materials Science and Engineering Division, of the
U.S. Department of Energy under Contract No. DE-AC02-05CH11231.
S.C. acknowledges support from a Samsung Foundation.
\newpage

\textbf{Supplements}

\textbf{Areal density of MIGS.} We give a simple estimate of the areal density of
MIGS. In a two-band tight-binding model \cite{sankey02}, the amplitude squared of the
evanescent solutions \cite{kohn59} close to the valence band edge has an
energy-dependent decay length $\beta(E) = 2[2m^*(E-E_{VBM})/
 \hbar^2]^{1/2}$, where $m^*$
is the electron effective mass and $E_{VBM}$ is the energy of the valence
band maximum. Near the conduction band edge $E_{CBM}$, $\beta(E) = 2[2m^*(E_{CBM} -
 E)/ \hbar^2]^{1/2}$. The areal density $N$ of MIGS in the insulator (in units of
states per unit area) is given by \cite{cohen76}
\begin{equation}\label{eq2}
N=\int_{0}^{E_F}dE\int_{0}^{\infty}dz\eta(E)e^{-\beta(E)z}=\eta\int_{0}^{E_F}dE\frac{1}{\beta(E)}
\end{equation}
where we have assumed the density of states $\eta(E)$ of the metal to be
constant over the energy range of the band gap. Inserting the
expression for $\beta(E)$ into Eq.(\ref{eq2}), we obtain
\begin{equation}\label{eq3}
N=\eta[(\hbar^2/2m^*)(E_F-E_{VBM})]^{1/2}
\end{equation}
For most semiconductors and insulators \cite{yu05}, $m_e/m^* \approx
1+C_1/E_g$ and $E_F-E_{VBM} = C_2E_g$ with $C_1 \approx 10$ eV and $C_2
\approx$ 0.5; furthermore, for most metals $\eta(E)$ is of the same of
order of magnitude. Consequently, the approximate expression
\begin{equation}\label{eq4}
N \approx \eta [(\hbar^2/2m_e)C_1C_2]^{1/2}
\end{equation}
is relatively insensitive to the nature of both the metal and the
insulator. Using the typical values $\eta(E) \approx
2\times10^{28}$m$^{-3}$eV$^{-1}$ and $C_1C_2 \approx 5$eV, we obtain $N \approx 8 \times 10^{18} \mathrm{m}^{-2}$, in good agreement with pseudopotential
calculation \cite{cohen77} for Al in contact with Si, GaAs or ZnS.

\textbf{Hubbard energy $U_i$.} We calculate the Hubbard energy $U_i$ for double
occupation for states in the isolated D region by evaluating the
integral
\begin{equation}\label{eq5}
U_i=\int_{D}d\mathbf{r}d\mathbf{r'}\frac{|\varphi_{i,\uparrow}(\mathbf{r})|^2|\varphi_{i,\downarrow}(\mathbf{r'})|^2}{|\mathbf{r}-\mathbf{r'}|}
\end{equation}
over the supercell. Within our tight-binding supercell scheme, two
additional factors need to be included. (i) The part of the Coulomb
integral on the same atomic site is replaced with the value of an onsite
Hubbard $U_0$. (ii) When the localization length ($\xi$) of the localized
states is larger than the supercell size, there is overlap of
wavefunctions from the neighboring supercell; this overestimates $U_i$ for
the very weakly localized states. Given that the participation number,
$P_i=1/\sum_j|\varphi_i(\textbf{r}_j)|^4 \sim (\xi_i/a)^d$
in a disordered $d$-dimensional system with supercell lattice constant $a$
and $U_i \propto 1/\xi_i$, we map the $U_i$ value of the finite supercell onto that of an
infinite supercell using a scaling law \cite{kramer81} for $\xi$.

\textbf{Hybridization energy broadening $\mathit{\Gamma_i}$.} The hybridization-energy broadening
of the localized states arises from couplings to the extended states in
the metal as well as those in the insulator, and is given by 
\begin{equation}\label{eq6}
\mathit{\Gamma_i}=\mathit{\Gamma_i^M}+\mathit{\Gamma_i^I}
\end{equation}
\begin{equation}\label{eq7}
\mathit{\Gamma_i^M}=\pi|V_i^M|_{ave}^2\rho^M(E),\mathit{\Gamma_i^I}=\pi|V_i^I|_{ave}^2\rho^I(E)
\end{equation}
where $\rho^{M(I)}(E)$ is the density of extended states in M (I) at the energy of the
localized state $E$, and $V_i^{M(I)}$ is the hopping matrix element between an
extended state in M(I) and a localized state in D ($ave$ indicates
averaging over the extended states). Extended eigenstates in M(I) are a
linear combination of constituent orbitals; the $V_i^{M(I)}$ can then be expressed
in terms of the coupling of these orbitals to those in D. For example,
the localized states inside the band gap of the insulator are hybridized
with only extended states in M, and $\mathit{\Gamma_i}=\mathit{\Gamma_i^M}\approx\pi
V^2d_i/W$. (Here $d_i$ is the charge of the
localized state $\mid i\rangle$ in the unit cell layer immediately adjacent to M.)


\newpage
\begin{figure}
\includegraphics*[width=80mm]{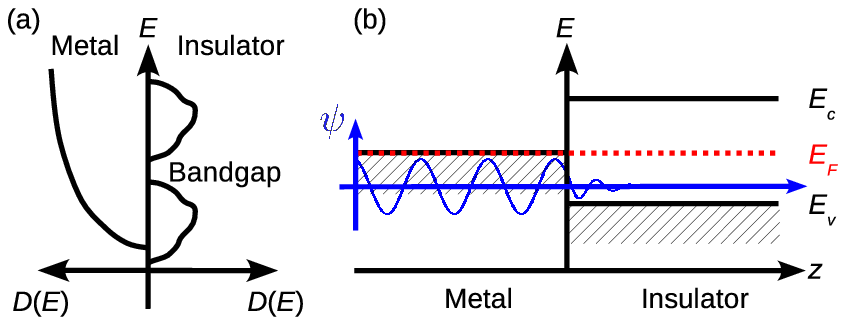}
\caption{\label{fig1} (Color online) 
(a) Schematic density of states. (b) MIGS at a
perfect interface with energy in the band gap are extended in the metal
and evanescent in the insulator.}
\end{figure}

\begin{figure}
\includegraphics*[scale=1.0]{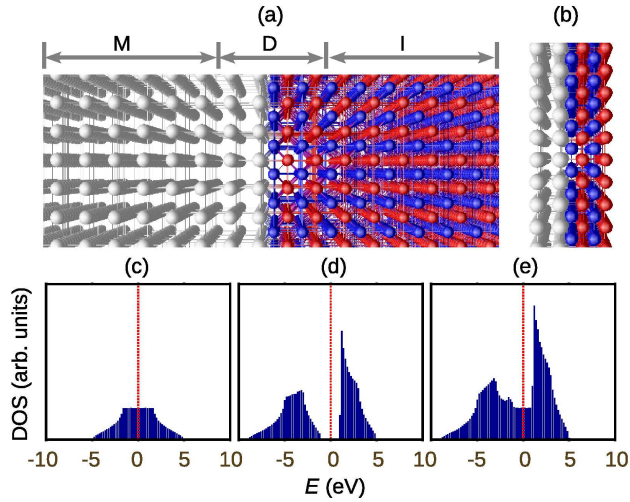}
\caption{\label{fig2} (Color online) 
(a) The metal (M) has a simple cubic structure with one
atom per unit cell and the insulator (I) a CsCl structure with two atoms
per unit cell. (b) Interfacial region (D) consists of 2 layers of metal
unit cells and 2 layers of insulator unit cells. The lattice constant is
0.15 nm. Computed DOS with Fermi energy (dotted red line)
set to zero. (c) Typical metal with 10 eV bandwidth. (d) Typical insulator
with a 2 eV band gap separating two bands of about 8 eV and 4 eV. (e)
Metal-insulator interface with MIGS in the band gap of the insulator
due to the presence of the metal.}
\end{figure}

\begin{figure}
\includegraphics*[scale=1.0]{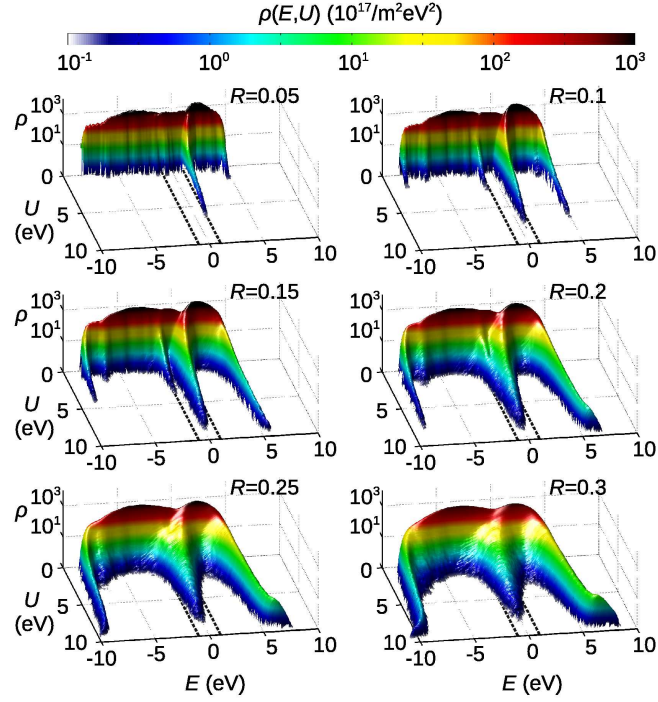}
\caption{\label{fig3} (Color online) 
Density of states distribution $\rho(E,U)$ as a function of energy $E$ and Hubbard
energy $U$ for 6 values of the randomness parameter $R$ in the isolated D
region of Fig. 2. For a given value of $R$, the highest values of $U$,
resulting in the most highly localized states, appear in the band gap of
the insulator and at the band edges. The position of the insulator band gap is represented by black dashed lines.}
\end{figure}

\begin{figure}
\includegraphics*[scale=1.0]{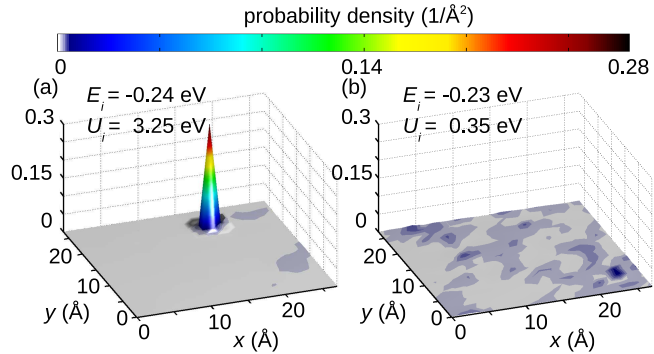}
\caption{\label{fig4} (Color online) Perspective view images of the
 2-dimensional probability density distribution at the interfacial
 region (D) along directions parallel to the interface (x- and
 y-directions), integrated along the z-direction. (a) States with
$U_i$=3.25 eV and $E_i$=-0.24eV and (b) with $U_i$=0.35 eV and
$E_i$=-0.23eV, respectively.}
\end{figure}

\begin{figure}
\includegraphics*[scale=1.0]{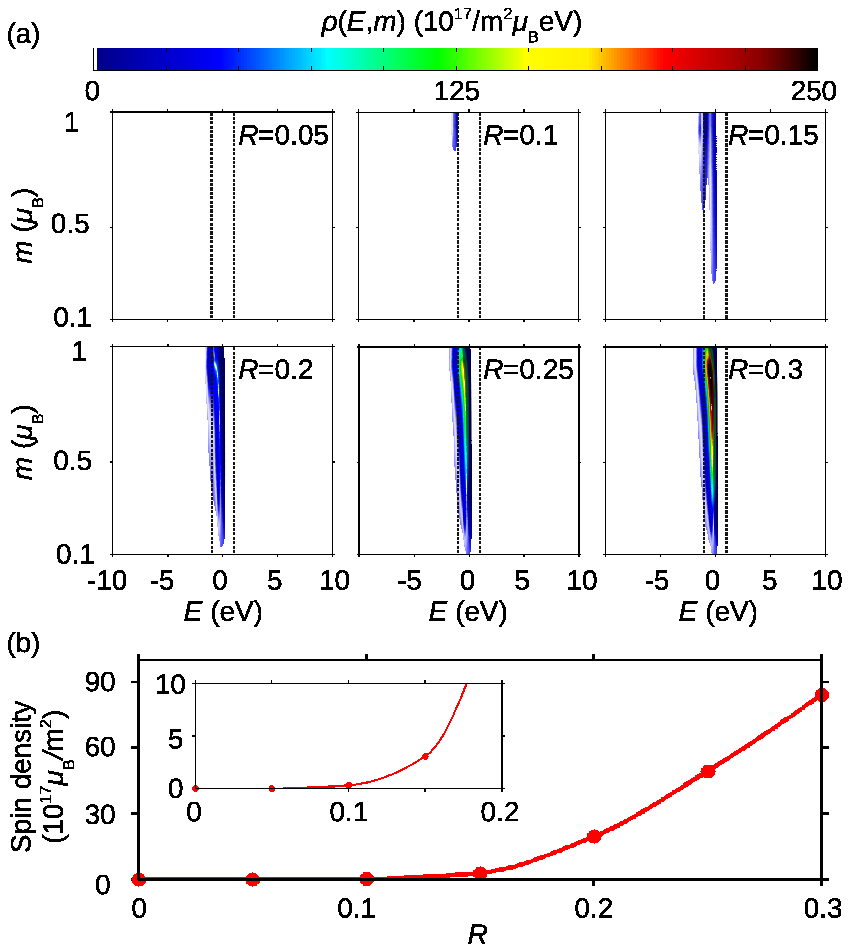}
\caption{\label{fig5} (Color online) 
(a) Electron density distribution $\rho(E,m)$ for 6 values of $R$. We simulated 5000 different
configurations of disorder for each value of $R$. The position of the
insulator band gap is represented by black dashed lines. Virtually all
the magnetic moments are from the MIGS in the band gap of the
insulator. (b) Integrated spin density versus randomness parameter
$R$. For $R=0.05$, we estimate the spin density to be less than $0.01
\times 10^{17}m^{-2}$.}
\end{figure}


\newpage
\begin{thebibliography}{natbib}
\bibitem{clarke04}J.Clarke and A.I.Braginski, \textit{The SQUID Handbook} (Wiley-VCH,
 GmbH and Weinheim, 2004), Vol. 1.
\bibitem{clarke87}F.C.Wellstood, C.Urbina, and J.Clarke, Appl. Phys. Lett. \textbf{50},
 772 (1987).
\bibitem{yoshihara06}F.Yoshihara, K.Harrabi, A.O.Niskanen, Y.Nakamura, and J.S.Tsai, Phys. Rev. Lett. \textbf{97}, 167001 (2006).
\bibitem{kakuyanagi07}K.Kakuyanagi \textit{et al.}, Phys. Rev. Lett. \textbf{98}, 047004 (2007).
\bibitem{bialczak07}R.C.Bialczak \textit{et al.}, Phys. Rev. Lett. \textbf{99}, 187006 (2007).
\bibitem{lanting09}T.Lanting \textit{et al.}, Phys. Rev. B \textbf{79}, 060509(R) (2009).
\bibitem{clarke07}R.H.Koch, D.P.DiVincenzo, and J.Clarke, Phys. Rev. Lett. \textbf{98},
267003 (2007). 
\bibitem{machlup54}S.J.Machlup, J. Appl. Phys. \textbf{25}, 341 (1954).
\bibitem{dutta81}P.Dutta and P.M.Horn, Rev. Mod. Phys. \textbf{53}, 497 (1981).
\bibitem{desousa07}R.de Sousa, Phys. Rev. B \textbf{76}, 245306 (2007).
\bibitem{faoro08}L.Faoro and L.B.Ioffe, Phys. Rev. Lett. \textbf{100}, 227005 (2008).
\bibitem{sendelbach08}S.Sendelbach \textit{et al.}, Phys. Rev. Lett. \textbf{100}, 227006 (2008).
\bibitem{bluhm_ar}H.Bluhm, J.A.Bert, N.C.Koshnick, M.E.Huber, and K.A.Moler, Phys. Rev. Lett. \textbf{103}, 026805 (2009).
\bibitem{cohen76}S.G.Louie and M.L.Cohen, Phys. Rev. B \textbf{13}, 2461 (1976). 
\bibitem{nyquist28}H.Nyquist, Phys. Rev. \textbf{32}, 110 (1928).
\bibitem{louie75}M.L.Cohen, M.Schlüter, J.R.Chelikowsky, and S.G.Louie,
Phys. Rev. B \textbf{12},5575 (1975).
\bibitem{cohen77}S.G.Louie, J.R.Chelikowsky, and M.L.Cohen, Phys. Rev. B \textbf{15},
2154 (1977). 
\bibitem{anderson58}P.W.Anderson, Phys. Rev. \textbf{109}, 1492 (1958).
\bibitem{anderson75}P.W.Anderson, Phys. Rev. \textbf{124}, 41 (1961).
\bibitem{epaps}See supplements
\bibitem{screening}If one includes the effect of metallic screening from
  region M on $U_i$ (Ref. \cite{cohen08}), $U_i$ would decrease by a
  factor of roughly 2 since the localized
  state in region I is located on average $\sim 3$ unit cell layers from region
  M. We estimate this effect reduces the spin density by $\sim 50\%$ at
  each $R$ value. As a result, $R$ should be incresed by at most $10\%$ to
produce an areal density of $\sim 5 \times 10^{17}m^{-2}$.
\bibitem{kittel54}M.A.Ruderman and C.Kittel, Phys. Rev. \textbf{96}, 99 (1954).
\bibitem{kasuya56}T.Kasuya, Prog. Theor. Phys. \textbf{16}, 45 (1956).
\bibitem{yosida57}K.Yosida, Phys. Rev. \textbf{106}, 893 (1957).
\bibitem{weissman93}M.B.Weissman, Rev. Mod. Phys. \textbf{65}, 829 (1993).
\bibitem{harris08}R.Harris \textit{et al.}, Phys. Rev. Lett. \textbf{101}, 117003 (2008).
\bibitem{fetter71}A.L.Fetter and J.D.Walecka, \textit{Quantum Theory of Many-Particle
Systems}, (McGraw-Hill, New York, 1971) p.298.
\bibitem{sendelbach_p}S.Sendelbach, D.Hover, M.M\"{u}ck, and
  R.McDermott, Phys. Rev. Lett. \textbf{103}, 117001 (2009).
\bibitem{cohen08}J.D.Sau, J.B.Neaton, H.J.Choi, S.G.Louie, and
  M.L.Cohen, Phys. Rev. Lett. \textbf{101}, 026804 (2008).
\bibitem{sankey02}J.K.Tomfohr and O.F.Sankey, Phys. Rev. B \textbf{65}, 245105 (2002). 
\bibitem{kohn59}W.Kohn, Phys. Rev. \textbf{115}, 809 (1959).
\bibitem{yu05}P.Y.Yu and M.Cardona, \textit{Fundamentals of Semiconductors: Physics and
Materials Properties} (Springer, Berlin, 2005). 
\bibitem{kramer81}A.MacKinnon and B.Kramer,
  Phys. Rev. Lett. \textbf{47}, 1546 (1981). 
\end{thebibliography}
\end{document}